\documentclass[doublecol]{epl2}
\usepackage{amsmath,amssymb}
\usepackage{graphicx,soul}

\newcommand{\dd}{\mathrm{d}}
\newcommand{\td}[2]{\frac{\dd #1}{\dd #2}}
\newcommand{\pd}[2]{\frac{\partial #1}{\partial #2}}
\newcommand{\mean}[1]{\langle #1 \rangle}
\newcommand{\Int}[1]{\int\dd #1\;}

\renewcommand{\vec}[1]{\mathbf #1}

\newcommand{\kap}{\kappa}
\newcommand{\lam}{\lambda}
\newcommand{\vhi}{\varphi}

\newcommand{\id}{\mathbf 1}

\newcommand{\x}{\vec r}
\newcommand{\kT}{k_\text{B}T}
\newcommand{\nois}{\boldsymbol\xi}
\newcommand{\Da}{D_\text{a}}
\renewcommand{\tx}{\tau_\text{r}}
\newcommand{\mmu}{\boldsymbol\mu}

\begin{document}

\title{Active Brownian particles driven by constant affinity}

\author{Thomas Speck}
\institute{Institut f\"ur Physik, Johannes Gutenberg-Universit\"at Mainz,
  Staudingerweg 7-9, 55128 Mainz, Germany}

\abstract{Experimental realizations of self-propelled colloidal Janus particles exploit the conversion of free energy into directed motion. One route are phoretic mechanisms that can be modeled schematically as the interconversion of two chemical species. Here we consider the situation when the difference of chemical potential between the two species (the driving \emph{affinity}) can be assumed to be constant, and we derive the thermodynamically consistent equations of motion. In contrast to the standard model of active Brownian particles parametrized by a constant self-propulsion speed, this yields a non-constant speed that depends on the potential energy of the suspension. This approach allows to consistently model the breaking of detailed balance and the accompanying entropy production without non-conservative forces.}


\maketitle

%% ---- introduction ----

\section{Introduction}

At thermal equilibrium, the microscopic dynamics governing the motion of particles obeys \emph{detailed balance}, a condition that guarantees the absence of directed transport (no preferred direction, no currents) and a vanishing entropy production. The fundamental symmetry is time-reversal: in equilibrium we cannot distinguish whether a movie is played forward or backward. This is different in driven systems, with the dissipation rate determining time asymmetry~\cite{feng08} and bounding uncertainties~\cite{bara15,ging16}. Consequently, understanding how detailed balance is broken in driven systems is pivotal for consistent and accurate modeling.

Locally driven \emph{active particles} (both cellular~\cite{pros15} and colloidal~\cite{bech16}) that undergo directed motion have received enormous attention recently. The particles are characterized by an orientation along which they are propelled, and that evolves in time. Even if the average orientation is zero implying the absence of particle currents (no mass transport), the dynamics breaks detailed balance with a non-vanishing heat dissipation. In experiments on active colloidal particles, the energy to sustain the directed motion is supplied locally, typically through light~\cite{jian10,butt12} or chemically through the decomposition of hydrogen peroxide~\cite{paxt04,hows07,pala10}. Such active particles have become the focus of intensive research due to, among many other reasons, novel collective behavior like motility induced phase separation in the absence of attractive forces~\cite{thom11,yaou12,butt13,bial14,cate15} and possible applications in the self-assembly of colloidal materials~\cite{kumm15,meer16}. Active suspensions have already been exploited to power microscale devices~\cite{soko09,leon10,kais14,krish16} and for templated self-assembly~\cite{sten16}.

For simple models of active particles the entropy production has been studied but with conflicting definitions and results~\cite{gang13,fala16a,fodo16,spec16,mand17,marc17}. The reason is that these models describe the effective motion of the particles but ignore the physical mechanism that leads to the self-propulsion. Only very recently this mechano-chemical coupling has been taken explicitly into account for active particles~\cite{gasp17,piet17}. In this Letter, we derive the governing equations of motion for Janus particles propelled through the conversion of two chemical species at fixed chemical potential. In response to every chemical event, the particle is translated by a distance $\lam$. In the limit of small $\lam$, we recover the equations employed for active Brownian particles but with a non-constant speed that depends on the change of potential energy as dictated by thermodynamic consistency. We calculate the work for a single active particle trapped in a harmonic potential numerically and analytically, and demonstrate that, as the frequency of chemical events is increased, the work approaches the limit behavior obtained for constant speed.

\section{Active Brownian particles}

A minimal model for active particles is active Brownian particles (ABPs)~\cite{bial12}. The positions obey
\begin{equation}
  \label{eq:lang}
  \dot\x_k = -\mu_0\nabla_k U + v_0\vec e_k + \nois_k,
\end{equation}
where $U(\{\x_k\})$ is the potential energy due to repulsive conservative forces between particles and the second term describes the self-propulsion along the particle orientation $\vec e_k$ with constant speed $v_0$. The Gaussian white noises have zero mean and correlations $\mean{\nois_k(t)\nois^T_l(s)}=2D_0\delta_{kl}\mathbf 1\delta(t-s)$ with strength $D_0=\kT\mu_0$, where $T$ is the temperature of the solvent (acting as the heat bath), $\mu_0$ is the bare Stokes mobility, and $\beta=(\kT)^{-1}$ with Boltzmann's constant $k_\text{B}$. The orientations undergo free rotational diffusion.

ABPs resolve the positions $\{\x_k\}$ and orientations $\{\vec e_k\}$ of particles. They do not resolve the underlying mechanism of the self-propulsion. Consequently, it is not obvious what the behavior of the self-propulsion under time reversal is, leading to two possibilities: (i)~it does not change sign corresponding to a non-conservative force or (ii)~it does change sign. We will show that the latter is the case when taking the conversion of (free) energy explicitly into account. The entropy production rate $\dot{\mathcal S}$ (as calculated from time reversal, see appendix) of ABPs then reads
\begin{equation}
  \dot{\mathcal S}/\beta
  = -\sum_{k=1}^N\left(\dot\x_k-v_0\vec e_k\right)\cdot\nabla_kU
  \overset{!}{=} \dot q.
\end{equation}
The stochastic action of the orientations is symmetric and does not contribute to the entropy production. Thermodynamic consistency requires to identify the dissipated heat $\dot q$ with the entropy produced in the heat bath. From the first law $\td{U}{t}=\dot w-\dot q$ we thus obtain the work rate
\begin{equation}
  \label{eq:w:abp}
  \dot w_\text{ABP} = \sum_{k=1}^Nv_0\vec e_k\cdot\nabla_kU
\end{equation}
spent externally to maintain the non-equilibrium steady state. This expression is in agreement with general considerations on the invariance of work and heat with respect to the frame of reference~\cite{spec08}.

%% ---- motor ----

\section{Interlude: Molecular motor}

\begin{figure}[t]
  \centering
  \includegraphics{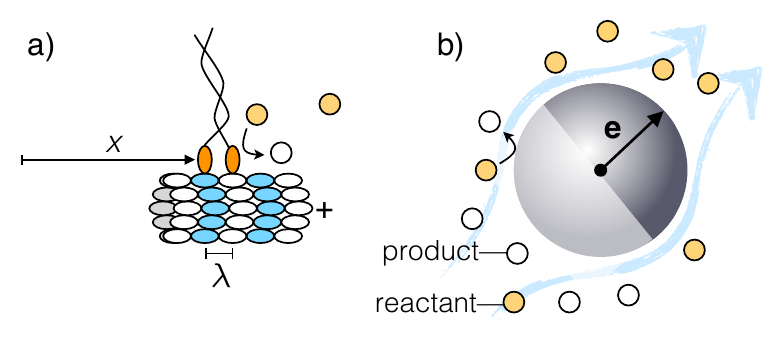}
  \caption{Mechano-chemical coupling. (a)~Sketch of a molecular motor moving along a microtubule and loaded with constant force $F$. Each directed step with step size $\lam$ is due to the conversion of an ATP molecule ($\bullet\to\circ$). (b)~Colloidal Janus particle with two different hemispheres defining its orientation $\vec e$. One hemisphere catalyzes an enzymatic reaction, which depletes the reactants close to this hemisphere. This creates a gradient, which causes a hydrodynamic slip velocity $\vec u$ propelling the particle. We assume that every reaction causes a discrete ``jump'' of the particle along $\vec e$ with step length $\lam$.}
  \label{fig:motor}
\end{figure}

Consider a single molecular motor (\emph{e.g.} kinesin) moving along a microtubule [Fig.~\ref{fig:motor}(a)]. In equilibrium, the motion is diffusive with zero mean displacement. The motor can be described as an enzyme to which reactant (substrate) molecules $\bullet$ (typically ATP--adenosine triphosphate) bind, hydrolysis of which induces a conformal change that leads to a directed step (with the direction determined by the polarity of the microtubule), after which the product $\circ$ (ADP and P$_\text{i}$) is released. We denote this reaction $\bullet\rightleftharpoons\circ$ with chemical potential difference $\Delta\mu\equiv\mu_\bullet-\mu_\circ$. In addition, the motor performs useful work through lifting a weight. Note that tight coupling simplifies the description, but more realistic models with several internal states also fall into our framework. These models include idle cycles~\cite{liep07}, during which the reservoir work is dissipated without producing useful work.

A thermodynamically consistent set of transition rates is obtained through combining the molecular motor with reactant and product molecules into a super-system~\cite{seif18,lee18}. This super-system is coupled to a heat bath at temperature $T$. At the coarsest level of description, we assume tight coupling so that every time a reactant molecule is bound and hydrolyzed the motor performs a step~\cite{juli97,kolo13}. In this limit, $n=\Delta n_\circ$ is the number of reactions and thus the number $\Delta n_\circ$ of product molecules produced during the observation time. The total number of reactant and product molecules $n_\bullet+n_\circ=n_\text{tot}$ is conserved. The work reservoir thus holds the Gibbs free energy
\begin{equation}
  \label{eq:G}
  G_\text{res}(n) = \mu_\bullet n_\bullet + \mu_\circ n_\circ = \mu_\bullet n_\text{tot} 
  - n\Delta\mu,
\end{equation}
which takes on the familiar bilinear form. Clearly, $n$ is an extensive quantity and $\Delta\mu$ corresponds to the thermodynamics force, also called affinity.

The total energy of the combined system plus reservoir is $E(x,n)=U(x)+G_\text{res}(n)$, where $U(x)$ is the potential energy of the system depending on the position $x$ of the motor along the microtubule. Chemical events are coupled to conformation changes, which implies that also $x$ changes. The rates for the chemical events obey detailed balance
\begin{equation}
  \label{eq:db}
  \frac{\kap^+}{\kap^-} = \frac{\kap(x\to x',n\to n+1)}{\kap(x'\to x,n+1\to n)} 
  = \frac{\Psi_\text{eq}(x',n+1)}{\Psi_\text{eq}(x,n)}
\end{equation}
with respect to the equilibrium distribution $\Psi_\text{eq}(x,n)\propto e^{-\beta E(x,n)}$. However, for $\Delta\mu\neq 0$ the combined system will not reach equilibrium but a stationary state in which the chemical reservoir is ``drained'', \emph{i.e.}, it performs work on the system (the molecular motor) that is eventually dissipated as heat. The first law of thermodynamics for a single chemical event reads
\begin{equation}
  \label{eq:firstlaw}
  \delta E = E(x',n+1)-E(x,n) = \delta U - \Delta\mu \overset{!}{=} -\delta q.
\end{equation}
The combined system only exchanges energy with the heat reservoir, which thus is heat $\delta q$. Consequently, we identify $\Delta\mu$ as the work with first law (from the perspective of the motor) $\delta U=\Delta\mu-\delta q$ and total work $w_\text{res}=n\Delta\mu$.

%% ---- derivation ----

\section{Self-propelled colloidal particles}

Instead of a molecular motor, let us now consider a solvated colloidal particle with an inhomogeneous surface, \emph{e.g.}, a spherical Janus particles with one hemisphere coated with a catalyst that promotes a chemical reaction [Fig.~\ref{fig:motor}(b)]. The simplest model is a generic chemical reaction $\bullet\rightleftharpoons\circ$ of a molecular solute~\cite{gole05,gole07}, quite in analogy with the molecular motor. The majority of experimental studies on self-propelled Janus particles exploits the decomposition of hydrogen peroxide (the molecular solute) into hydrogen and oxygen~\cite{bial14,bech16}. An alternative mechanism is the reversible demixing of the molecular solute (specifically, lutidine in water)~\cite{butt12}. In both cases, self-propulsion is powered by the difference in chemical potential (either between reactants and products, or the two phases). Quite generally, an imbalance of local fluxes and mobilities across the particle surface leads to a hydrodynamic slip velocity $\vec u=\lam\dot n\vec e$ due to the reciprocal theorem for Stokes flow~\cite{gole07}. Here, $\vec e$ is the unit orientation of the particle (for a Janus particle it points along the poles of the two hemispheres), $\dot n$ is the total flux of molecular solutes, and $\lam$ is a length that depends on the particle geometry and other specific factors. It is fairly challenging to actually calculate $\lam$, the most common approach being based on the thin boundary layer approximation~\cite{ande89}. For our purposes, however, it is sufficient to retain $\lam$ as a parameter.

We now return to a suspension of $N$ active particles. The total number of new product molecules is $\Delta n_\circ=n=\sum_kn_k$ with $n_k$ the number of reactions occurring on the surface of the $k$-th particle. The total number $n_\text{tot}=n_\bullet+n_\circ$ of solute molecules remains constant with the Gibbs free energy of the work reservoir given by Eq.~\eqref{eq:G}. The change of potential energy due to a \emph{single} chemical event reads
\begin{equation}
  \delta U = U(\x_k+\lam\vec e_k) - U(\x_k) \approx \lam\vec e_k\cdot\nabla_kU
  \equiv \hat f_k
\end{equation}
holding all other positions fixed. Here we have assumed that $\lam\ll\ell$ is much smaller than the typical length $\ell$ set by the interactions (typically related to the particle size). The detailed balance condition Eq.~\eqref{eq:db} then reads
\begin{equation}
  \label{eq:db:k}
  \frac{\kap^+_k}{\kap^-_k} = e^{-\beta(\hat f_k-\Delta\mu)},
\end{equation}
where the right hand side only depends on the positions $\{\x_k\}$ plus the orientation of the $k$th particle. For the rates, it is thus sufficient to indicate the particle positions as argument with $\kap^+_k(\x_k)=\kap^+_k(\x_k\to\x_k+\lam\vec e_k)$ and $\kap^-_k(\x_k)=\kap^-_k(\x_k+\lam\vec e_k\to\x_k)$.

Between the discrete reactions, the system evolves according to dynamics obeying detailed balance with respect to the total energy $E$, and thus the potential energy $U$. Hence, the evolution equation $\partial_t\Psi=\mathcal L_\text{p}\Psi+\mathcal L_\text{a}\Psi$ for the joint probability $\Psi(\{\x_k,\vec e_k,n_k\};t)$ can be split into a ``passive'' contribution with differential operator
\begin{equation}
  \mathcal L_\text{p} = \sum_{k=1}^N 
  \nabla_k\cdot\left[\mu_0(\nabla_kU)+D_0\nabla_k\right] + \mathcal L_\text{rot}
\end{equation}
with $\mathcal L_\text{rot}$ governing the rotations, and the active translations due to the chemical reactions. Assuming the chemical events to occur independently, the active operator is given by $\mathcal L_\text{a}\Psi = \sum_{k=1}^N \mathcal L_k\Psi$ with
\begin{multline}
  \label{eq:La}
  \mathcal L_k\Psi = \kap^+_k(\x_k-\lam\vec e_k)\Psi(\x_k-\lam\vec e_k,n_k-1) 
  \\ + \kap^-_k(\x_k)\Psi(\x_k+\lam\vec e_k,n_k+1)
  \\ -  [\kap^+_k(\x_k)+\kap^-_k(\x_k-\lam\vec e_k)]\Psi(\x_k,n_k),
\end{multline}
where, for clarity, we only indicate the arguments that are affected. For the first two terms, we expand the joint probability $\Psi(\x_k\pm\lam\vec e_k,n_k\pm1)$ as
\begin{equation}
  \Psi\pm\lam\vec e_k\cdot\nabla_k \Psi 
  + \frac{1}{2}\lam\vec e_k\cdot\nabla_k\pd{\Psi}{n_k} \pm\pd{\Psi}{n_k}+\cdots
\end{equation}
treating the $n_k$ as real numbers for simplicity. In a second step, we also expand the rates $\kap^\pm_k(\x_k-\lam\vec e_k)$. We then sum over the numbers $n_k$ of chemical events with the marginal joint probability
\begin{equation}
  \psi(\{\x_k,\vec e_k\};t) \equiv \left[\prod_{k=1}^N\Int{n_k}\right]
      \Psi(\{\x_k,\vec e_k,n_k\};t).
\end{equation}
To linear order of $\lam$, we thus obtain the operators
\begin{equation}
  \label{eq:La:approx}
  \mathcal L_k\psi \approx
  -\nabla_k\cdot[\lam(\kap^+_k-\kap^-_k)\vec e_k\psi]
\end{equation}
describing the propulsion of the particles with speeds
\begin{equation}
  \label{eq:v}
  \hat v_k(\{\x_l\},\vec e_k) \equiv \lam(\kap^+_k-\kap^-_k)
\end{equation}
that depend on particle positions.

\begin{figure*}
  \centering
  \includegraphics{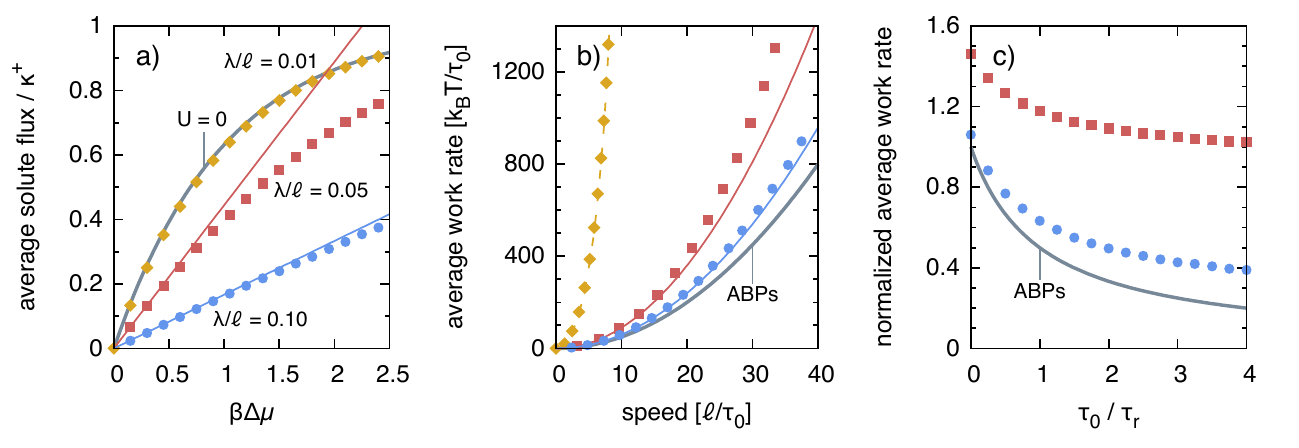}
  \caption{Numerical results for the combined dynamics of particle and reactions $n$ in a harmonic trap. (a)~Average solute flux $\mean{\dot n}$ as a function of $\Delta\mu$ for three lengths $\lam/\ell$ for $\tau_0/\tx=1$ and $\kap^+\tau_0=10^3$. Symbols are numerical results. The solid lines show the linear behavior from Eq.~(\ref{eq:ndot:lin}). The thick gray line shows the limiting behavior in the absence of a potential ($U=0$). (b)~The average work rate $\mean{\dot w_\text{res}}=\mean{\dot n}\Delta\mu$ as a function of speed $\mean{\hat v}=\lam\mean{\dot n}$. The solid lines show the quadratic behavior Eq.~(\ref{eq:wres:lin}) for small speeds. The dashed line is the limiting behavior for $U=0$ with maximal speed $v_\infty=\lam\kap^+$. The thick gray line shows the work rate Eq.~(\ref{eq:w:cf}) for ABPs. (c)~Average work rate as function of inverse orientation time $\tau_0/\tx$ and normalized by $\mean{\hat v}^2\tau_0/(\beta\ell^2)$. The thick gray line shows the limiting behavior $(1+\tau_0/\tx)^{-1}$.}
  \label{fig:num}
\end{figure*}

The corresponding Langevin equations
\begin{equation}
  \label{eq:abp:v2}
  \dot\x_k = \hat v_k\vec e_k - \mu_0\nabla_kU + \nois_k
\end{equation}
are our central result (they agree with Ref.~\cite{piet17} but lack an active noise, which is of order $\lam^2$). They describe the evolution of the particle positions without explicit reference to the chemical events, which are typically impossible to resolve in experiments. However, even if we only observe positions and orientations, the underlying propulsion mechanism due to the chemical events implies a non-constant speed $\hat v_k$ in contrast to the standard model of active Brownian particles [Eq.~\eqref{eq:lang}]. While we cannot fully resolve the stochastic reservoir work $w_\text{res}$ anymore, we can still determine its average $\mean{\dot w_\text{res}}=\mean{\dot n}\Delta\mu$. We find
\begin{align}
  \mean{\dot w_\text{res}} &= 
  \sum_{k=1}^N \mean{(\kap^+_k-\kap^-_k)(\Delta\mu-\hat f_k+\hat f_k)} \\
  &\approx \sum_{k=1}^N\left[\mean{\hat v_k\vec e_k\cdot\nabla_kU} 
  + \beta\mean{\kap^+_k(\hat f_k-\Delta\mu)^2} \right]
\end{align}
after exploiting the detailed balance condition Eq.~\eqref{eq:db:k} and expanding the exponential to first order in the second term. The first contribution is the work corresponding to the trajectory entropy, cf. Eq.~\eqref{eq:w:abp}, but with $\hat v_k$ instead of $v_0$. However, there is an additional contribution which manifestly stems from the fluctuations of $\hat f_k$ away from the prescribed driving affinity $\Delta\mu$.

%% ---- numerics ----

\section{Illustration: Harmonic trap}

To illustrate the consequences of taking into account the chemical reactions, we now turn to a single active particle moving in two dimensions in the external harmonic potential $U=\tfrac{1}{2}k\x^2$ with $\hat f=k\lam(\vec e\cdot\x)$. The stiffness $k$ sets a natural length $\ell\equiv(\beta k)^{-1/2}$ and time scale $\tau_0\equiv(\mu_0k)^{-1}$. The unit orientation $\vec e=(\cos\vhi,\sin\vhi)^T$ is expressed by the angle $\vhi$ it encloses with the $x$-axis. We assume that this orientation undergoes free rotational diffusion with correlation time $\tx$, which leads to the passive evolution operator
\begin{equation}
  \mathcal L_\text{p}\psi = \frac{1}{\tau_0}\nabla\cdot(\x\psi) 
  + D_0\nabla^2\psi + \frac{1}{\tx}\pd{^2\psi}{\vhi^2}.
\end{equation}
The time evolution of the average $\mean{\hat f}$ involving $\psi$ can be written
\begin{equation}
  \partial_t\mean{\hat f} = 
  -\left(\frac{1}{\tau_0}+\frac{1}{\tx}\right)\mean{\hat f} 
  + k\lam^2\mean{\dot n}
\end{equation}
after inserting the evolution equation with Eq.~(\ref{eq:La:approx}), and performing integrations by part with vanishing boundary terms. Since we are interested in the steady state, we set the time derivative on the left hand side to zero. Note that we have a choice for the rates $\kap^\pm$ as long as they obey the detailed balance condition Eq.~(\ref{eq:db:k}). Here, we assume that the rate $\kap^+$ is a constant. After expanding $\mean{\dot n}\approx\kap^+\beta\mean{\Delta\mu-\hat f}$ we solve for $\mean{\hat f}$ and finally obtain the expression
\begin{equation}
  \label{eq:ndot:lin}
  \mean{\dot n} \approx 
  \frac{\kap^+\beta\Delta\mu}{1+(\lam/\ell)^2\kap^+\tau_0/(1+\tau_0/\tx)}
\end{equation}
valid in the linear regime $\mean{\dot n}\propto\beta\Delta\mu$ of small driving affinity. In the same linear regime, for the average reservoir work rate $\mean{\dot w_\text{res}}=\mean{\dot n}\Delta\mu$ we obtain
\begin{equation}
  \label{eq:wres:lin}
  \mean{\dot w_\text{res}} = 
  \left(\frac{1}{1+\tau_0/\tx}+\frac{1}{\kap^+\tau_0(\lam/\ell)^2}\right)
  \frac{\mean{\hat v}^2\tau_0}{\beta\ell^2}
\end{equation}
with average propulsion speed $\mean{\hat v}=\lam\mean{\dot n}$. For comparison, the average work rate Eq.~\eqref{eq:w:abp} for ABPs reads
\begin{equation}
  \label{eq:w:cf}
  \mean{\dot w_\text{ABP}} = \frac{1}{1+\tau_0/\tx}\frac{v_0^2\tau_0}{\beta\ell^2}
\end{equation}
for all speeds $v_0$. Hence, the average work spent by the reservoir is always larger than forcing a constant speed $v_0$ without fluctuations.

Using a kinetic Monte Carlo scheme for the transitions $n\leftrightharpoons n+1$ in addition to integrating the discretized Langevin equations, we have solved numerically the full stochastic dynamics of the particle \emph{and} the reactions. In the following we set $\kap^+\tau_0=10^3$ and the integration time step to $\Delta t=10^{-3}\tau_0$. The result for the average solute flux is plotted in Fig.~\ref{fig:num}(a) for three values of $\lam/\ell$. Also shown is the limiting result $\mean{\dot n}_0=\kap^+(1-e^{-\beta\Delta\mu})$ in the absence of a potential, which is approached for $\lam/\ell\to0$ (since the potential difference for directed steps becomes negligible). We see that Eq.~(\ref{eq:ndot:lin}) indeed describes the linear regime for small $\Delta\mu$. The range of validity of the linear approximation increases as $\lam/\ell$ becomes larger. In Fig.~\ref{fig:num}(b) we plot the corresponding work rate, which for sufficiently large $\lam/\ell$ and smaller speeds is well approximated by the quadratic expression Eq.~(\ref{eq:wres:lin}). Increasing $\lam/\ell$ further, the work rate approaches that of ABPs with constant propulsion speed $v_0$. Both expressions for the average work become equivalent in the limit
\begin{equation}
  \label{eq:equiv}
  \kap^+\tau_0 \gg \frac{1+\tau_0/\tx}{(\lam/\ell)^2}
\end{equation}
of large solute flux and, consequently, small fluctuations. Note that in the opposite limit corresponding to a vanishing external potential the work Eq.~(\ref{eq:wres:lin}) is determined by the first term alone, and thus the model of ABPs is no longer applicable. In Fig.~\ref{fig:num}(c), we show the average work changing the orientational correlation time $\tx$.

Choosing for the particle radius $a=\ell=1\,\mu$m, one obtains $\tau_0\approx10\,$s for water at room temperature. Hence, speeds on the order of $\mu$m/s are reached for driving affinities $\Delta\mu$ of a few $\kT$ consuming 100 reactant molecules per second. These speeds agree with what is observed for self-propulsion due to the demixing of a near-critical binary water-lutidine solvent~\cite{butt12,butt13}.

%% ---- excess work ----

\section{Neglecting translational noise}

As observed in computer simulations, the translational noise on the particle positions has little influence on the large-scale behavior, in particular one still observes a motility induced phase separation~\cite{yaou12}. This has motivated a modification of ABPs with
\begin{equation}
  \label{eq:aoup}
  \dot\x_k = -\mu_0\nabla_kU + \vec u_k, \qquad
  \tx\dot{\vec u}_k = -\vec u_k + \nois_k
\end{equation}
called the active Ornstein-Uhlenbeck process (AOUP)~\cite{fodo16}. The noise now stems from the fluctuations of the orientations $\vec u_k$ ($\nois_k$ is Gaussian with noise strength $\Da>0$), which are not normalized anymore. The orientational correlations are still determined by $\tx$.

Conceptually, the limit $D_0\to0$ would imply $T\to0$ of the heat bath. There are two options to proceed: one can interpret Eq.~(\ref{eq:aoup}) as equations of motion arising from some non-equilibrium medium and construct thermodynamic notions in analogy to stochastic thermodynamics. This route has been followed in Refs.~\cite{fodo16,mand17} for the AOUP (see also Refs.~\cite{gang13,pugl17} for similar treatments). Both works map the coupled equations of motion to an underdamped model for which they calculate the path entropy following the standard approach of stochastic thermodynamics. These works arrive at different expressions and conclusions, which highlights the conceptual difficulties of this route. In particular, Ref.~\cite{fodo16} posits a continuation of the effective equilibrium regime to linear order of $\tx$ with vanishing path entropy production at variance with established results for the linear response regime. Moreover, even for a harmonic potential the authors predict a vanishing entropy production. Ref.~\cite{mand17} posits that an additional term besides the dissipated heat is required to restore the second law. Such a modification of the second law is not plausible for the physical mechanism underlying the directed motion and, as shown here, not necessary.

The arguably more transparent route is, for the same \emph{physical} system, to interpret these equations as effective equations of motion neglecting the translational noise. The influence of the heat bath now only enters through the dynamics of $\vec u_k$. Replacing $v_0\vec e_k\to\vec u_k$, the expressions for work and heat thus remain unchanged, in particular Eq.~(\ref{eq:w:abp}) is the work on the particles. For a single particle moving in the harmonic potential, we now obtain $\mean{\vec u\cdot\x}=\Da/(1+\tx/\tau_0)$. Choosing $\Da=v_0^2\tx$, we recover exactly the same work rate Eq.~(\ref{eq:w:cf}) as for active Brownian particles with constant speed $v_0$.

If, instead, we control the noise strength $\Da$ and orientational correlation time $\tx$ independently as suggested in Ref.~\cite{fodo16}, we obtain
\begin{equation}
  \label{eq:w:aoup}
  \mean{\dot w_\text{alt}} = \frac{1}{1+\tx/\tau_0}\frac{\Da}{\beta\ell^2}.
\end{equation}
In Ref.~\cite{fodo16} it has been shown that in the limit $\tx\to0$ the stationary distribution $\Psi\propto e^{-\beta_\text{eff}U}$ approaches a Boltzmann distribution at an effective temperature $\kT_\text{eff}=\Da/\mu_0$. However, Eq.~(\ref{eq:w:aoup}) shows that the work and thus the dissipation do not vanish in this limit. The behavior is thus fundamentally different from active colloidal particles [cf. Fig.~\ref{fig:num}(c)], which for $\tx\to0$ reach thermal equilibrium with vanishing dissipation.

%% ---- excess work ----

\section{Excess work}

In Ref.~\cite{spec16}, we have explored the idea that dissipation of ABPs can be modeled as an effective non-conservative force $\vec f_k=-(v_0/\mu_0)\vec e_k$. While here we have shown that the dissipation has to be modeled as a time-asymmetric ``flow'' term in Eq.~(\ref{eq:abp:v2}) due to the underlying coupling to chemical reactions, the expression for the excess work (perturbing the non-equilibrium steady state) remains the same in both approaches. To this end, we insert the Langevin equation~\eqref{eq:lang} into the work Eq.~\eqref{eq:w:abp},
\begin{equation}
  \dot w_\text{ABP} = \sum_{k=1}^N v_0\vec e_k\cdot\nabla_kU
  = \sum_{k=1}^N \vec f_k\cdot[\dot\x_k-v_0\vec e_k-\nois_k].
\end{equation}
Perturbing the particle positions $\{\x_k\}$ through an external parameter $X$ thus requires the excess work
\begin{equation}
  \delta w_\text{ex} = \left[\pd{E}{X} 
    + \sum_{k=1}^N\vec f_k\cdot\pd{\x_k}{X}\right]\delta X
\end{equation}
with an additional term due to the work required to keep the system in the non-equilibrium steady state. Hence, all conclusions of Ref.~\cite{spec16} regarding the pressure and interfacial tension of ABPs remain valid for the identification of work and heat proposed here. The consequences of a non-constant speed $\hat v_k$ for the pressure remain to be studied.

%% ---- conclusions ----

\section{Conclusions}

We have studied a schematic mechanism in which particles are translated in response to each conversion of a molecular solute driven by a non-zero chemical potential difference $\Delta\mu$ (the driving affinity). This mechanism yields a thermodynamically consistent modification of active Brownian particles [Eq.~(\ref{eq:abp:v2})], in which the propulsion speed now depends on the change of the potential energy. The consequences of this modification for the collective dynamics and the motility induced phase transition will be explored elsewhere.

In this first step, we have assumed a constant driving affinity and neglected a spatial dependence of the concentration of molecular solutes driving the propulsion, assuming a ``pervading'' reservoir of reactant molecules. In more realistic situations, however, these molecules might only be exchanged at the system's boundary. Consumption of molecules on the particle surfaces then induces depletion and long-range concentration profiles, giving rise to phoretic interactions between the colloidal particles. Several numerical schemes have been introduced to explicitly take into account the concentration profile of the molecular solute~\cite{thak11,yan16a,huang17}. The model presented here [Eq.~(\ref{eq:abp:v2})] is intermediate between these schemes and active Brownian particles with a constant speed. It is suitable to study the emergent collective behavior in systems coupled to a large reservoir keeping the driving affinity constant. Moreover, we have treated the solvent as a structureless ideal medium, whereas in a real fluid the solvated colloidal particles will induce correlations and currents. Both effects could be included in the theory presented here on the level of a Gaussian field theory~\cite{spec13}.

While assuming a strictly constant driving affinity is an approximation, at least some experiments have been performed under conditions that come close to this assumption. For example, in Ref.~\cite{pala10} experiments on Janus particles driven by the decomposition of hydrogen peroxide are performed in a chamber molded into agarose gel. Through the gel (away from the chamber) a solution with constant concentration of hydrogen peroxide is circulated. This ensures the renewal of fuel and removal of the chemical waste products by diffusion through the gel walls. The system is thus coupled to an almost ideal reservoir with constant concentration, which determines the chemical potential difference.

% outlook
To study the collective behavior of active matter, typically coarse-grained dynamic equations are employed~\cite{nard17}. To ensure consistency with the microscopic heat dissipation, novel algorithms to systematically construct such coarser models from the microscopic equations of motion are needed. Progress in this direction has been made recently through a cycle-based approach~\cite{knoc15}. Finally, the concept of reservoirs naturally introduces intensive variables out of equilibrium~\cite{bertin06}, which might pave the way to novel numerical algorithms (``grand-canonical'' simulations with fluctuating particle number~\cite{meer17}) and help to further rationalize non-equilibrium phase coexistence~\cite{taka15a,solo17,paliwal18}.

%% ---- acknowledgments ----

\acknowledgments

I thank Michael E. Cates and Udo Seifert for illuminating and helpful discussions. Useful discussions during a visit of the International Centre for Theoretical Sciences (ICTS) participating in the program ``Stochastic Thermodynamics, Active Matter and Driven Systems'' (Code: ICTS/Prog-stads2017/2017/08) are acknowledged. The DFG is acknowledged for financial support within priority program SPP 1726 (grant number SP 1382/3-2). Part of this work has been supported by the Humboldt foundation through a Feodor Lynen alumni sponsorship.

%% ---- appendix ----

\section{Appendix: Time reversal}

The stochastic action corresponding to Eq.~(\ref{eq:lang}) reads
\begin{equation}
  \label{eq:act:A}
  \mathcal A = \Int{t} \frac{1}{4D_0}\sum_{k=1}^N
  \left(\dot\x_k-v_0\vec e_k+\mu_0\nabla_kU\right)^2
  + \mathcal A_\text{rot},
\end{equation}
where $\mathcal A_\text{rot}$ is the action due to the rotational motion. Depending on stochastic calculus there are additional terms, which, however, are irrelevant for the entropy production. Denoting time reversal by $\mathcal A^\dagger$ mapping $\dot\x_k\mapsto-\dot\x_k$ and $v_0\vec e_k\to-v_0\vec e_k$, the part of the action that is asymmetric under time reversal is identified with the (dimensionless) entropy production
\begin{equation}
  \label{eq:act:S}
  \mathcal S = \mathcal A^\dagger - \mathcal A 
  = -\beta\Int{t}\sum_{k=1}^N(\nabla_kU)\cdot(\dot\x_k-v_0\vec e_k) = \beta q,
\end{equation}
which equals the heat $q$ dissipated into the heat bath at inverse temperature $\beta$.

Specifically, for two-dimensional free rotational diffusion we have
\begin{equation}
    \mathcal A_\text{rot} = \frac{1}{4}\tx\sum_{k=1}^N \dot\vhi_k^2,
\end{equation}
which is manifestly symmetric, $\mathcal A_\text{rot}^\dagger=\mathcal A_\text{rot}$, for all times.

Including hydrodynamic coupling between the active Brownian particles, the Langevin equation~(\ref{eq:lang}) for the reference positions becomes
\begin{equation}
  \delta\vec v_k \equiv \dot\x_k - v_0\vec e_k 
  + \sum_{l=1}^N\mmu_{kl}\cdot\nabla_lU = \nois_k,
\end{equation}
where the symmetric mobility matrices $\mmu_{kl}$ depend on particle separations. The same mobility matrices now determine the noise correlations
\begin{equation}
  \label{eq:hyd}
  \mean{\nois_k(t)\nois_l^T(t')} = 2\kT\mmu_{kl}\delta(t-t')
\end{equation}
so that the stochastic action reads
\begin{equation}
  \mathcal A = \frac{\beta}{4}\Int{t} \sum_{k,l=1}^N
  \delta\vec v_k\cdot\mmu^{-1}_{kl}\cdot\delta\vec v_l
\end{equation}
with $\sum_i\mmu_{ki}\cdot\mmu^{-1}_{il}=\delta_{kl}\id$. Calculating the asymmetric contribution Eq.~(\ref{eq:act:S}), we find the same result as in the absence of hydrodynamic interactions. This demonstrates that, as long as Eq.~(\ref{eq:hyd}) is fulfilled, the dissipation along a \emph{single trajectory} is not influenced by the hydrodynamic coupling, see also Ref.~\cite{spec08}.

%% ---- bibliography ----

\end{document}